\documentclass[aps,prl,twocolumn,nofootinbib]{revtex4-1}
\usepackage{amsmath}
\usepackage{amsfonts}
\usepackage{amssymb}
\usepackage{color}
\usepackage{bm}		% bolded Greek letters
\usepackage{graphicx}
\usepackage{feynmp}
\usepackage{cancel}
\usepackage{slashed}	% for slashed characters in math mode
\usepackage{bbm}		% for \mathbbm{1} (unit matrix)
\usepackage{xspace}	% For spacing after commands
\usepackage{empheq}
\usepackage[unicode]{hyperref}	% Has to be at the end.
\usepackage[utf8]{inputenc}

\providecommand{\be}{\begin{equation}} 
\providecommand{\ee}{\end{equation}}

\def\Br{\mathop{\mathrm{Br}}}
\newcommand\un[1]{\,\mathrm{#1}} % recommended in order to prevent bad line-breaking.
\newcommand\MeV{\un{MeV}}
\newcommand\GeV{\un{GeV}}
\newcommand\TeV{\un{TeV}}
\newcommand\fb{\un{fb}}

\newcommand\TO{\text{--}}

\def\QCD{\mathrm{QCD}} % SM
\providecommand{\oln[1]}{\overline{#1}} % Overline, MS-bar scheme, appendix B
\def\a{\alpha}

\def\g{\gamma}

\def\j{\psi}

\def\q{\theta}

\def\s{\sigma}
\def\t{\tau}

\def\G{\Gamma}

\def\L{\Lambda}

\def\U{\Upsilon}

\newcommand{\w}[1]{_{\rm #1}}

\makeatletter
\protected\def\EE{\@ifnextchar-{\@@EE}{\@EE}}
\protected\def\@EE#1{\ifnum#1=1 \!\times\!10 \else \!\times\!10^{#1}\fi}
\protected\def\@@EE#1#2{\!\times\!10^{-#2}}
\makeatother

\newcommand{\secetaX}{\texorpdfstring{$\boldsymbol{\eta_X}$}{\texteta\_X}}
\newcommand{\secXXbar}{\texorpdfstring{$\boldsymbol{X}$--$\boldsymbol{\bar X}$}{X--Xbar}}
\newcommand{\secLambdaH}{\texorpdfstring{$\boldsymbol{\Lambda_h}$}{\textLambda\_h}}

\begin{document}

%----------------------------------------------------------------------------------------

\title{Diphoton Signals from Colorless Hidden Quarkonia}
\author{Sho Iwamoto}
\email{sho@physics.technion.ac.il}
\author{Gabriel Lee}
\email{leeg@physics.technion.ac.il}
\author{Yael Shadmi}
\email{yshadmi@physics.technion.ac.il}

\affiliation{Physics Department, Technion---Israel Institute of Technology,\\ Haifa 32000, Israel}

\author{Robert Ziegler}
\email{robert.ziegler@lpthe.jussieu.fr}
\affiliation{Sorbonne Universit\'es, UPMC Univ Paris 06, UMR 7589, LPTHE, F--75005, Paris, France\\
       CNRS, UMR 7589, LPTHE, F--75005, Paris, France}

%----------------------------------------------------------------------------------------

\date{\today}
\begin{abstract}
We show that quarkonia-like states of a hidden SU($N$) gauge group
can account for the $750\GeV$ diphoton excess observed by ATLAS and CMS,
even with constituents carrying standard model hypercharge only.
The required hypercharge is modest, varying between about 1.3--1.6 for strong
SU($N$) coupling, to 2--3 for weak SU($N$) coupling, for $N=3,4$.
This scenario predicts a variety of diphoton and multi-photon resonances,
as well as photons from continuum pair production, and possibly exotic
decays into standard model fermions, with no significant multi-jet resonances.
\end{abstract}

\maketitle

%----------------------------------------------------------------------------------------

\section{Introduction}

%----------------------------------------------------------------------------------------

If new particles are produced at the LHC, they have so far eluded detection,
suggesting some suppression of their decays. 
In the presence of such suppression, bound states of these particles can be produced near 
threshold.  Diphoton resonances from  bound-state decay may then be the first 
harbingers of new physics.
In this paper, we interpret the excess of diphoton events near $750\GeV$ reported by
the ATLAS~\cite{ATLAS-CONF-2015-081, ATLAS-CONF-2016-018, Aaboud:2016tru} 
and CMS~\cite{CMS:2015dxe, CMS:2016owr, Khachatryan:2016hje} Collaborations 
as arising from the decay of a ``quarkonium''-like state, $\eta_X$, 
bound by a hidden confining SU($N$).
Our model is minimal in that it assumes photoproduction as the main
$\eta_X$ production channel~\cite{Fichet:2015vvy,Csaki:2015vek,Csaki:2016raa}.
Thus, we take the $\eta_X$ constituents $X$ to carry hypercharge only, 
but no standard model (SM) SU(3)${}_c$ or SU(2)${}_L$ quantum numbers.
This is in contrast to similar recent work which featured bound states 
with colored constituents~\cite{Harigaya:2015ezk,Nakai:2015ptz,Curtin:2015jcv,Agrawal:2015dbf,Han:2016pab,Harigaya:2016pnu,Kats:2016kuz,Harigaya:2016eol,Kamenik:2016izk,Ko:2016sht}.
While the production cross section is controlled by the hypercharge of the 
constituents, $Y_X$, 
and is proportional to $Y_X^4$, bound-state formation is controlled by the 
new strong force.

We will focus on the possibility that the constituents $X$ are vector-like fermions 
in the (anti-)fundamental representation of the hidden SU($N$).%
\footnote{Scenarios with scalar constituents are possible as well, leading to different phenomenology.
For example, the DY production of $J=1$ state is suppressed due to the lack of spin.}
The lowest-lying $J = 0$ bound state is mainly produced via vector-boson fusion (VBF) processes, 
especially $\gamma\gamma$ fusion, while Drell--Yan (DY) production gives $J=1$ bound states.

Below we discuss two limiting cases, for which the bound-state properties
can be readily estimated.
In the first, the SU($N$) is weakly coupled, such that the SU($N$) confinement 
scale $\Lambda_h$ is much smaller than the constituent mass $m_X$.
The bound state is governed by the Coulomb SU($N$) potential in this case, 
so that its properties and production cross section can be calculated perturbatively.
In the second, the SU($N$) is strongly coupled, and 
confinement effects are important. 
We will then rely on analogies with measured QCD quantities,
mainly in the charmonium system, to infer
the properties of $\eta_X$.
For simplicity, we restrict our attention to high values of $\Lambda_h$,
such that $\eta_X$ decays into hidden glueballs are kinematically
forbidden.
We stress that intermediate values of $\Lambda_h$ can easily account for the 
diphoton signal, but we cannot give any quantitative estimate of
branching ratios, binding energies etc.\ in this region.

In either case, with no light SU($N$) flavors, 
the  models obtained are quirk-like models~\cite{ Kang:2008ea,Cheung:2008ke,
Curtin:2015jcv, Agrawal:2015dbf, Craig:2015lra} with uncolored quirks
and relatively large $\Lambda_h$, which leads to microscopic strings.
The distinguishing signatures of the models are a variety of multi-photon signatures,
with dijet or multi-jet resonances absent or suppressed.
In addition to diphoton events from bound-state decays,
$X$--$\bar X$ pair production above threshold also gives multi-photon final states, 
either directly from annihilations into photons, or from annihilations into hidden 
glueballs which in turn decay into photons (or SM fermions).

We also consider models with additional light SU($N$) flavors, which are SM-singlets, 
with couplings to $X$ and SM fields. 
$X$--$\bar X$ pair production above threshold then gives a pair of $X$-hadrons 
which decay to SM fields through those couplings.

\begin{figure}
\includegraphics[scale=0.6]{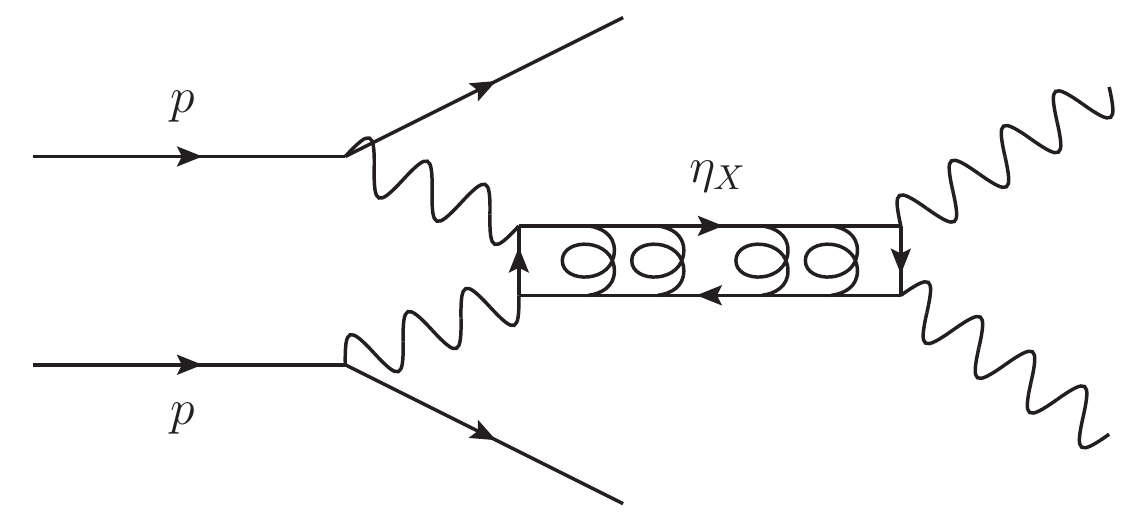}
\caption{Photoproduction of the bound state $\eta_X$ at the LHC. Gluon lines are for the hidden sector SU($N$).}
\label{fig:photoprod}
\end{figure}

This paper is organized as follows.
We begin by reviewing the required cross section to diphotons and decay rates for the bound states. 
We describe two scenarios, roughly corresponding to small and large hidden couplings, 
in which diphoton decays of the lightest $X\bar{X}$ bound state in the hidden sector can yield the correct signal.
We discuss the phenomenology of the hidden sector, including the possibility of excited bound states 
and LHC signals from pair production of $X$--$\bar{X}$.
Possible decays of $X$ itself are considered before we conclude.

\begin{table}[t]
\caption{Summary of the hidden SU($N$) and the SM charges of the new vector-like fermion $X$.}
\label{tab:charges}

\centering
\begin{tabular}{c|c|c}
& SU($N$) & SU(3)${}_c$, SU(2)${}_L$, U(1)${}_Y$ \\
\hline
$X$ & $N$ & $(1, 1, Y_X)$ \\
$\bar{X}$ & $\bar{N}$ & $(1, 1, -Y_X)$
\end{tabular}
\end{table}

%----------------------------------------------------------------------------------------

\section{The \secetaX{} bound state and diphoton signal strength}

%----------------------------------------------------------------------------------------

We begin with an overview of the different scales in the hidden sector.
For concreteness, we assume that the $750\GeV$ resonance is a $1^1S_0$ bound state $\eta_X$
of a fundamental Dirac fermion $X$ of mass $m_X$ and hypercharge $Y_X$ 
(cf.\ Table~\ref{tab:charges}).%
\footnote{Quarkonia are often labelled in the form $n_r{}^{2S+1} L_J$ in analogy with spectroscopic notation,
where $S$, $L$, and $J$ are the spin, orbital, and total angular momentum quantum numbers.
The radial excitation number $n_r$ is related to the principal quantum number $n$ by $n = n_r + L$. 
The radial excitation number and orbital angular momentum are used in parentheses for 
particle-like names of quarkonia states, e.g., $J/\j(1S)$.}
The required cross section is~\cite{Kats:2016kuz}
\be
\s(pp \rightarrow \eta_X \rightarrow \g\g) \Big|_{\sqrt{s} = 13\TeV} \sim 3\text{--}6\fb \,.
\ee
Since its constituents are not colored, the dominant production channel for $\eta_X$ is VBF,
specifically photon fusion. Taking the result from Ref.~\cite{Csaki:2016raa}, 
which includes the contributions from inelastic--inelastic, elastic--inelastic, and elastic--elastic processes, 
the total photoproduction signal strength at $13\TeV$ in the narrow width approximation is given by\footnote{The cross-section at $8\TeV$ is about a factor of~2 smaller, and thus in tension with Run~1 diphoton searches. However, this ratio is subject to potentially large uncertainties~\cite{Csaki:2015vek, Fichet:2015vvy}.} 
\begin{align} \label{eqn:xs}
\sigma_{13}^{p p \to \eta_X \to \gamma \gamma} & = 
5 \fb \left( \frac{\Gamma_{\mathrm{tot}}}{21 \MeV}\right) \Br(\eta_X \to \gamma \gamma)^2  \, ,
\end{align}
where $\Gamma_{\mathrm{tot}}$ is the total decay width of $\eta_X$.
 $Z\g$ and $ZZ$ production channels contribute an additional 8\% to the cross section~\cite{Kamenik:2016tuv}.

Bound-state production is enhanced for $L=0$ states by the wavefunction at the origin.
The partial width of the lightest bound state $\eta_X$ into photons is~(cf. Ref.~\cite{Kamenik:2016izk})
\be \label{eqn:decaygamgam}
\frac{\G(\eta_X \rightarrow \g\g)}M = 4 N (Y_X^2 \a)^2 \frac{|R_{n0}(0)|^2}{M^3} \,,
\ee
where $R_{n0}(r)$ is the radial wavefunction of a bound state with
orbital angular momentum $L = 0$ and principal quantum number $n$,
normalized such that $\int_0^\infty |R_{n0}(r)|^2 r^2 dr = 1$.
Perturbatively, the ratio of $\eta_X$ decay rates to two photons vs.\ two hidden gluons is
\be\label{ghgh}
\frac{\G(\eta_X \to \g\g)}{\G(\eta_X \to g_hg_h)} = 4\, \frac{N^2}{N^2 - 1}\, \frac{(Y_X^2 \a)^2}{\a_h^2} \,,
\ee
where $g_h$ denotes a hidden gluon.
Below, two scenarios are considered: 
one with small $\alpha_h$ such that the decay rates to diphotons and invisible hadrons are comparable,
and one with large $\alpha_h$, with the hidden glueball channel kinematically closed. 
We refer to these scenarios as ``Low~$\Lambda_h$" and ``High~$\Lambda_h$".

%----------------------------------------------------------------------------------------

\subsection{Low \secLambdaH}

For small $\alpha_h$, the SU($N$) binding potential can be described in the Coulomb approximation~\cite{Kats:2012ym}
\be
\label{eqn:pertwavefunc}
|R_{n0}(0)|^2 = \frac{C^3 \oln{\a}_h^3 m_X^3}{2n^3} \,,
\ee
where 
\be
C = \frac12 (C_1 + C_2 - C_\eta) \,,
\ee
and $C_\eta, C_1, C_2$ are the quadratic Casimirs of the bound state and its constituent particles.
For constituents in the (anti-)fundamental representation and the singlet bound state of $\bar{X} X$,
$C_{\eta} = 0$ and $C_1 = C_2 = \frac{N^2 - 1}{2N}$. 
Then
\be
|R_{n0}(0)|^2 \sim \left( \frac{N^2 - 1}{2N} \right)^3 \frac{\oln{\a}_h^3 M^3}{16n^3} \,,
\ee
where we assumed that the bound-state energy $M\sim 2m_X$, 
and defined $\oln{\a}_h$ as the hidden sector gauge coupling in the $\overline{\rm MS}$ scheme, 
evaluated at the Bohr radius of the bound state 
\begin{align}
r_{\text{rms}} \sim a_0 & = 2/(C\oln{\a}_h m_X),&
\overline{\alpha}_h & \equiv \alpha_h (a_0^{-1}) \, .
 \label{eqn:alphabardef}
\end{align}
The decay rate into photons is then given by
\be
\frac{\G(\eta_X \rightarrow \g\g)}M = \frac{1}{4}\, N\, Y_X^4\alpha^2\,
   \left( \frac{N^2 - 1}{2N} \right)^3 \oln{\a}_h^3\,,
\ee
where $\alpha = \alpha (M) \approx \alpha (M_Z) \approx 1/128.$

The binding energy is given by (at leading order)  
\be
E_n = -\frac1{4n^2} C^2 \oln{\a}_h^2 m_X \,,
\label{eqn:Eb}
\ee
so that the mass of this bound state is ($E_b \equiv E_1$)
\be
M = 2m_X + E_b = \bigg(2 - \frac14 \left( \frac{N^2 - 1}{2N} \right)^2  \oln{\a}_h^2 \bigg) m_X \,.
\label{eqn:BSmass}
\ee

%%%%%%%%%%%%%%%%%%%%%%%%%%%%%%%%%%%%%%%%%%%%%%%%%%%%%
\begin{figure}[t!]
\centering
\includegraphics[scale=0.55]{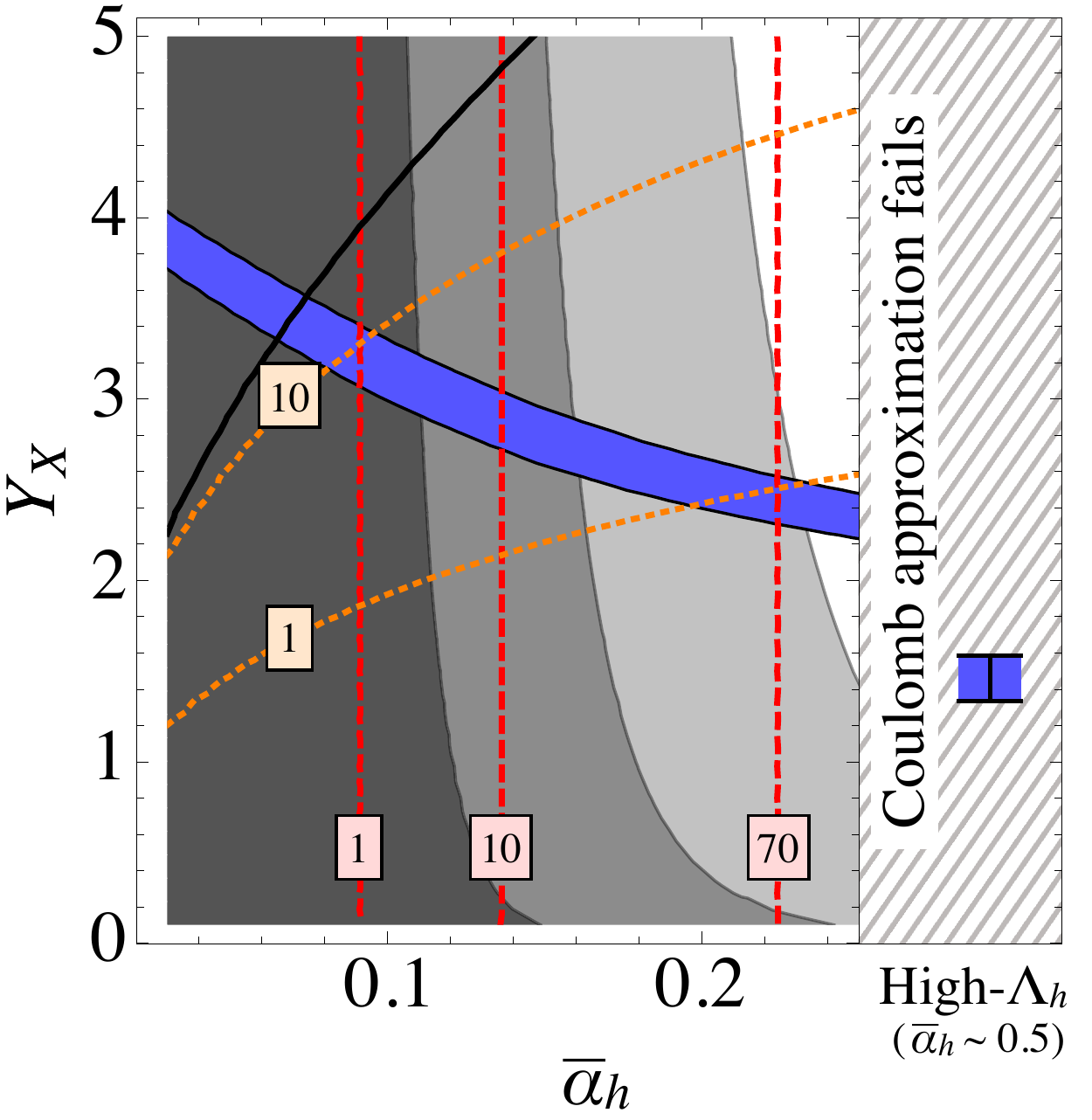}
\caption{Parameter space for $N=3$. The blue band indicates the diphoton signal strength of 3--$6\fb$, 
vertical red (dashed) lines denote the lightest glueball mass $M_{G_h}$ in GeV, 
and orange (dotted) lines denote the ratio $\Gamma(\eta_X \to \gamma\gamma)/ \Gamma(\eta_X \to g_h g_h)$. 
The grey-filled regions indicate the lifetime of the lightest glueball, ranging from left to right as 
  $\tau_{G_h} > 1 \un{s}$ (dark grey), 
  $1 \un{s} > \tau_{G_h} > 10^{-7} \un{s}$, 
  $10^{-7} \un{s} > \tau_{G_h} > 10^{-12} \un{s}$, and 
  $ \tau_{G_h} < 10^{-12} \un{s}$ (white). 
The black line indicates values of $\oln{\a}_h$ and $Y_X$ for which the binding energies from
hypercharge and the hidden SU$(N)$ are equal.
The value $\oln{\alpha}_h = 0.25$ corresponds to $\alpha_h(M) = 0.11$.
At larger $\oln{\alpha}_h$, we infer the signal strength from the QCD charmonium system 
and show it as a preferred range of $Y_X$.
See text for details.}
\label{fig:xsN3}
\end{figure}
%%%%%%%%%%%%%%%%%%%%%%%%%%%%%%%%%%%%%%%%%%%%%%%%%%%%%

To calculate the signal strength, we also need the partial decay rates 
into the hidden sector, and into different SM particles.
Since the resonance couples only to hypercharge, 
the decay rates into $Z\g$ and $ZZ$ are given by
\begin{align}
 \frac{\G(\eta_X \rightarrow \{ Z\g, ZZ \} )}{\G(\eta_X \rightarrow \g\g)}
 &= \{ 2\tan^2\q_W, \tan^4\q_W \} \nonumber \\
 & \approx \{0.6, 0.08\} \, , 
\end{align}
where $\q_W$ is the Weinberg angle with $\sin^2\q_W \approx 0.23$.
Including these channels reduces the branching ratio to diphotons by about 40\%.

Decays into hidden sector hadrons are dominated by annihilations into two hidden 
gluons, Eq.~\eqref{ghgh}. With no light hidden flavors, these hadronize into hidden glueballs. 
For pure QCD, the lightest glueball has $J^{PC} = 0^{++}$ and mass $\sim 7\Lambda_\QCD$~\cite{Morningstar:1999rf}.
We assume the same scaling for the lightest glueball $G_h$ of the hidden SU($N$),
$M_{G_h} \sim 7 \Lambda_h$,
where the confinement scale $\Lambda_h$ is given at one-loop order by
\begin{align}
 \label{lambdah}
  \Lambda_h & \sim m_X \exp\left( \frac{-2\pi}{b_0 \a_h(m_X)} \right) \,,
\end{align}
with $ b_0 = \frac{11}3 N - \frac23 N_F$, where $N_F$ is the number of light fermion flavors.
$G_h$ mainly decays to photons through loops of $X$,
with lifetimes estimated for $N=3$ (see, e.g., Refs.~\cite{Chen:2005mg, Juknevich:2009ji, Juknevich:2009gg}), 
\begin{align}\label{glueballtau}
\Gamma \left( 0^{++}_h \to \gamma \gamma \right) \approx \frac{Y_X^4 \alpha^2}{64 \pi^3} \frac{M_{G_h}^3}{m_X^2} \left( \frac{3 M_{G_h}^3}{60 m_X^3} \right)^2 \, .
\end{align}

%%%%%%%%%%%%%%%%%%%%%%%%%%%%%%%%%%%%%%%%%%%%%%%%%%%%%
\begin{figure}[t!]
\centering
\includegraphics[scale=0.56]{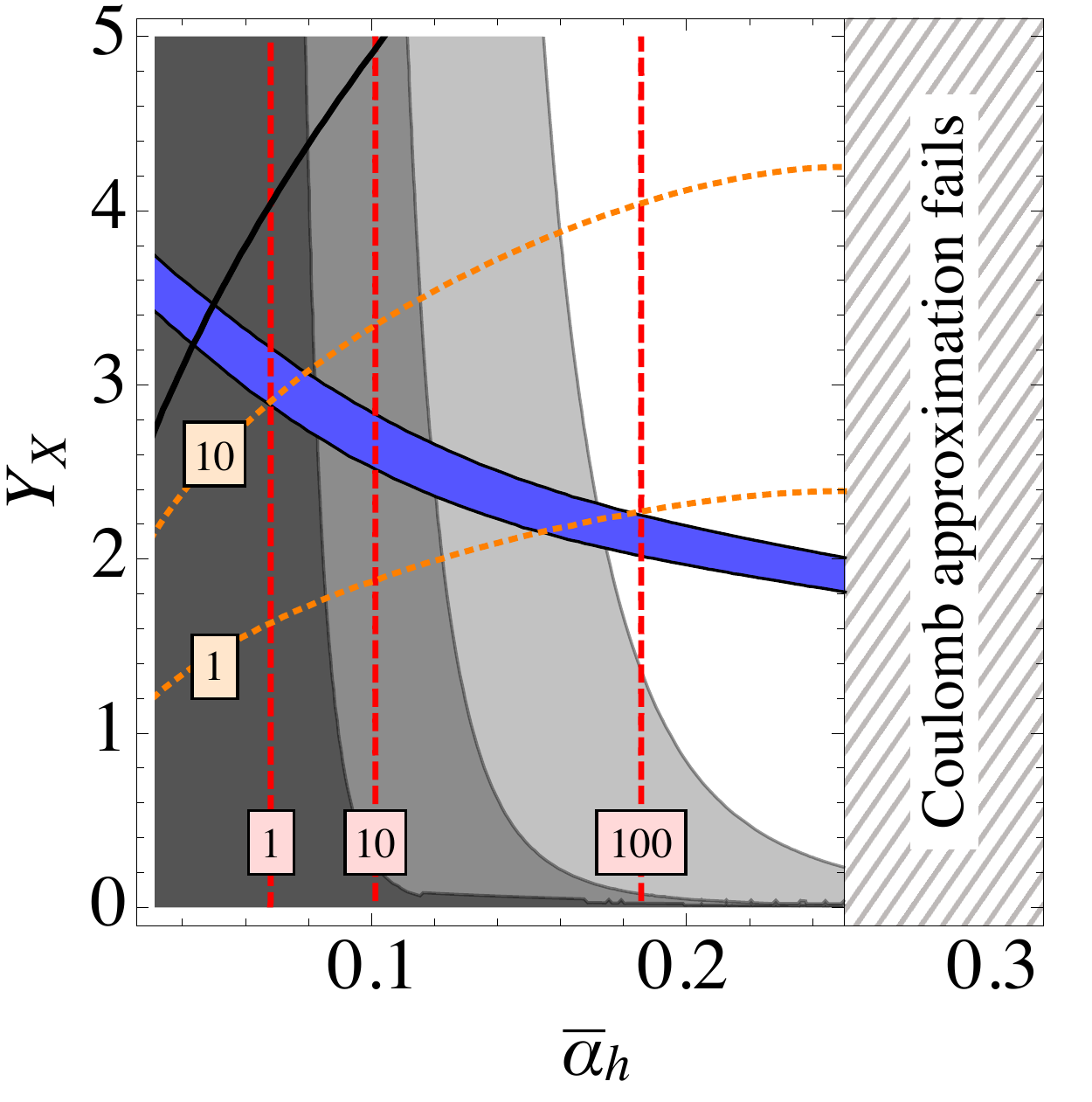}
\caption{Same as Fig.~\ref{fig:xsN3} for $N=4$. The value $\oln{\alpha}_h = 0.25$ corresponds to $\alpha_h(M) = 0.10$.}
\label{fig:xsN4}
\end{figure}
%%%%%%%%%%%%%%%%%%%%%%%%%%%%%%%%%%%%%%%%%%%%%%%%%%%%%

For fixed $N$, we are therefore left with $Y_X$ and $\oln{\a}_h$ as free parameters.
In Figs.~\ref{fig:xsN3} and \ref{fig:xsN4}, we show the required diphoton signal strength in the Coulomb regime
as a blue band in the $\oln{\alpha}_h$--$Y_X$ plane for $N=3$ and $N=4$, respectively, with $N_F=0$.
The plots are truncated at $\overline\alpha_h=0.25$, where the Coulomb approximation
can no longer be trusted (see discussion below).
Also indicated on the $N=3$ plot is the range of $Y_X$ preferred for the diphoton signal strength in the
``High-$\Lambda_h$'' scenario, which is discussed in the next subsection.
As expected, lower values of $Y_X$ are preferred in this case compared to the
Coulomb regime, since the enhancement factor $|R(0)|^2$ grows with the hidden coupling.
In between the two regions, phenomenological potentials,
which interpolate between the Coulomb and confining regimes, can be used to
describe the bound states (see, e.g., Ref.~\cite{Eichten:2007qx}).
Clearly, there are large uncertainties in these calculations.
Thus, for example, for stoponium,
a recent lattice calculation finds that potential 
models~\cite{Hagiwara:1990sq} underestimate 
$|R(0)|^2$ by a factor of about 4~\cite{Kim:2015zqa}.
In any case, the effect  of these modifications would be to increase the production cross section of $\eta_X$, 
so that the blue region would shift to smaller values of $Y_X$ above a given 
$\alpha_h$.

The vertical red (dashed) lines denote the lightest glueball mass in GeV; 
here, we determine $\Lambda_h$ by solving the two-loop renormalization group (RG) equations 
for the boundary condition $\a_h (\Lambda_h) = 4\pi$.
The orange (dotted) lines give the ratio $\Gamma(\eta_X \to \gamma\gamma)/ \Gamma(\eta_X \to g_h g_h)$ from Eq.~(\ref{ghgh}).
The grey-filled regions in Figs.~\ref{fig:xsN3} and \ref{fig:xsN4} indicate the lifetime of 
the lightest glueball according to Eq.~(\ref{glueballtau}), 
ranging from $>1\un{s}$ in the dark grey region to prompt decays in the white region.
The dark grey region is excluded by Big Bang nucleosynthesis (BBN) since 
energetic photons from glueball decays would dissociate nuclei.
The binding energy is less than $\mathcal{O}(10 \GeV)$ in this range;
we have included the contribution of the hypercharge Coulomb potential to the binding energy.%
\footnote{Including the hypercharge contribution in the binding energy and the wavefunction at the origin involves the substitution
$(C \oln{\a}_h)^n \to (C \oln{\a}_h + Y_X^2 \a)^n$ in Eqs.~(\ref{eqn:pertwavefunc}) and (\ref{eqn:Eb}).}
The solid black lines in the figures indicate values of $Y_X, \oln{\a}_h$ for which 
the contributions to the binding energy from hypercharge and the hidden SU$(N)$ are equal.
For $Y_X$ between 2--2.5, we obtain the diphoton signal for $\oln{\a}_h \gtrsim 0.22$ (0.18) 
for $N = 3$ $(4)$, which corresponds to $\alpha_h(M) \gtrsim 0.1$ $(0.08)$.

For very small values of $\alpha_h$, the cross section becomes independent of $\oln{\a}_h$
and the hypercharge Coulomb potential is dominant in creating the bound state.
The ratio of enhancement factors and binding energies are,
\be
\frac{|R(0)|^2_{\mathrm{em}}}{|R(0)|^2_{h}}
\sim \left(\frac{E_{b,\mathrm{em}}}{E_{b,\mathrm{h}}}\right)^{3/2}
\sim \left( \frac{2Y_X^2\alpha}{N\alpha_h} \right)^{3}\,.
\ee
Above the black lines in Figs.~\ref{fig:xsN3} and \ref{fig:xsN4}, the hypercharge Coulomb interaction is dominant.
In particular, we can use our results to estimate whether a purely hypercharged $X$ can account for the signal. 
This requires $Y_X \sim 4$, in contrast with larger values found in Ref.~\cite{Barrie:2016ndh} 
(taking into account the different multiplicity $N$ in our model).
Finally, we note that for $Y_X = 4$ and $N = 3$ $(N = 4)$, the hypercharge coupling $g_Y$ becomes non-perturbative 
at around 2000~(300)~TeV due to additional running from $X$.

For large $\oln{\alpha}_h$, we can understand the flattening of the blue signal region 
in Figs.~\ref{fig:xsN3} and \ref{fig:xsN4} from Eq.~(\ref{scaling}) as follows. 
It is instructive to consider how the total rate into photons scales with the SU($N$) parameters and $Y_X$.
From Eqs.~(\ref{eqn:xs}), (\ref{eqn:decaygamgam}), 
and (\ref{eqn:pertwavefunc}), and neglecting phase space factors due to the
hidden glueball mass, we have
\be\label{scaling}
\s(p p \to \eta_X \to \gamma \gamma) \propto Y_X^8 \left( \frac{\oln{\a}_h}{\a_h(M)} \right)^2 \oln{\a}_h \,,
\ee
which grows very fast with $Y_X$, and approximately linearly with the hidden coupling. 
In Eq.~(\ref{eqn:alphabardef}), we see that $a_0^{-1}$ grows linearly with $\oln{\alpha}_h$;
therefore, for large $\oln{\alpha}_h$, the hierarchy between $a_0^{-1}$ and $M$ itself is small,
so $\oln{\a}_h/\a_h(M)$ is approximately constant. 
Hence, to maintain a fixed cross section,
$Y_X$ needs to change only by a small amount to compensate for a given change in  $\overline{\alpha}_h$.

As $\alpha_h$ increases, confinement effects become important,
and the Coulomb approximation becomes inadequate.
Roughly speaking, this approximation is valid when
the Bohr radius of the bound state is larger than the confinement scale,%
\footnote{The potential approximation is valid for $\eta_b(1S)$, the lowest state of bottomonium, 
which has a separation scale $a_0 \sim 0.2$ fm compared to $\Lambda_\QCD^{-1} \sim 1$ fm \cite{Eichten:2007qx}.}
\be \label{eqn:pertcondLambda}
\Lambda_h \ll a_0^{-1} = \frac{C}2 \oln{\a}_h m_X = \left( \frac{N^2 - 1}{4N} \right) \oln{\a}_h m_X \,.
\ee
Furthermore, the radial wavefunction at the origin $|R(0)|^2$ and the binding energy 
as given in Eqs.~(\ref{eqn:pertwavefunc}) and (\ref{eqn:Eb}) are the leading-order results.
Higher-order corrections to the binding energy were calculated in~\cite{Penin:2002zv}. 
To obtain another estimate of the validity of the perturbative expansion, 
we take their result for the next-to-leading order (NLO) correction for zero light flavors,
\begin{align}
E_b & = E_1 \left[ 1 + \frac{\overline{\alpha}_h}{\pi} N  \times 2.85 + {\cal O} \left(\overline{\alpha}_h/\pi )^2 \right) \right] \, ,  
\end{align}
which implies that we need $\overline{\alpha}_h \lesssim 0.25$ in order to trust the perturbative expansion. 
We therefore truncate the plots at $\overline \alpha_h=0.25$.
For both Figs.~\ref{fig:xsN3} and \ref{fig:xsN4}, Eq.~(\ref{eqn:pertcondLambda}) is satisfied in this region.

The above discussion assumed no light SU($N$) flavors for concreteness,
but it can be simply extended to $N_F>0$ light flavors with mass of order
$\Lambda_h$. In this case, $\eta_X$ can decay to hidden hadrons as well,
whose mass is probably of order $\Lambda_h$, and lighter than the glueballs.
Furthermore,  $\Lambda_h$ is smaller for a given $\alpha_h$ because
of the slower running. For a single flavor, this is a mild effect, and the
results of Figs.~\ref{fig:xsN3} and \ref{fig:xsN4} will remain unchanged. In fact, the region in which the Coulomb approximation is valid will be wider in this case, allowing for larger
$\alpha_h$.

%----------------------------------------------------------------------------------------

\subsection{High \secLambdaH}

We now turn to consider larger SU($N$) couplings, for which the Coulomb approximation fails.
In this scenario, $|R(0)|^2$ (and hence the enhancement of bound-state production) and the binding energy 
are sensitive to confinement effects, and are therefore large.
In addition to the diphoton rates, $\eta_X$ will have large decay rates into hidden hadrons. 
Whether or not these are consistent with LHC constraints is hard
to estimate; so for simplicity, we focus here on very high $\Lambda_h$,
such that decays to glueballs are 
kinematically forbidden. 
In fact, a particularly dangerous channel is $\eta_X$ decay 
to a photon and an excited hidden glueball.
Using again the pure QCD results of Ref.~\cite{Morningstar:1999rf}, the lightest allowed glueball 
for this decay is the $1^{+-}_h$ glueball, with mass $\sim 1.7 M_{G_h}$.
Requiring this mass to be heavier than $\eta_X$, 
we have
\be
\Lambda_h \gtrsim 65 \GeV \,.
\ee
Thus, in this scenario, $\Br(\eta_X\to\gamma\gamma)\sim 0.6$.%
\footnote{Note that we cannot add light SU($N$) flavors in this case, 
since the resulting mesons would provide new, dangerous decay channels for
$\eta_X$.}

Since the Coulomb approximation cannot be used to compute bound-state quantities, 
 we must instead rely on analogies with measured QCD bound states, 
e.g., in the charmonium system (for a comprehensive review of quarkonia, see Ref.~\cite{Eichten:2007qx}).

%----------------------------------------------------------------------------------------

\begin{figure}
\includegraphics[scale=0.7]{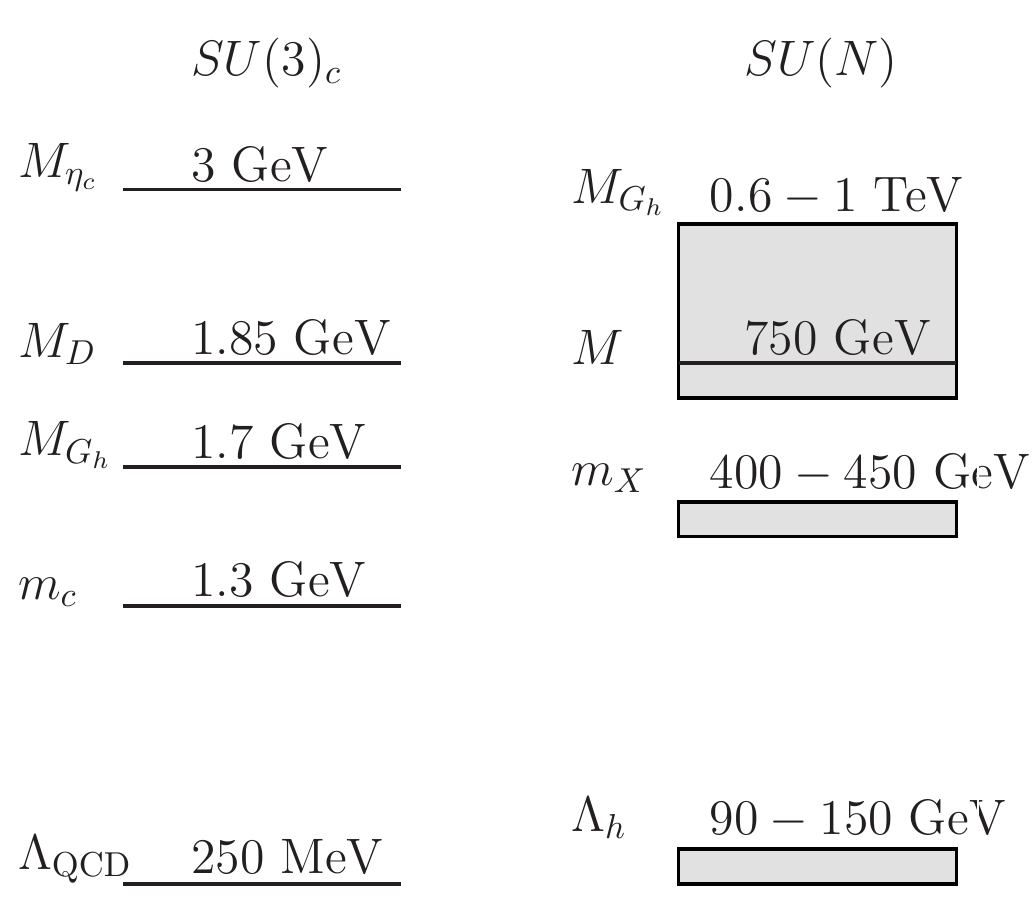}
\caption{Summary of scales for charmonium and QCD (left column) and for $\bar{X}X$ and the hidden SU($N$) (right column). 
Values of $\L_\QCD$ in the $\overline{\mathrm{MS}}$ scheme with various $N_F$ 
can be found in Ref.~\cite{Agashe:2014kda}; the FLAG review of Ref.~\cite{Aoki:2013ldr} quotes $\L_\QCD \sim 250\MeV$.
The values for the scales in the right column are representative, as detailed in the text, for an example with $N = 3$.}
\label{fig:largescales}
\end{figure}

\begin{table}[htb]
\centering
\caption{Summary of masses and widths for lowest states of charmonium and bottomonium, from Ref.~\cite{Agashe:2014kda}.}
\label{tab:qqmasses}
\begin{tabular}{c|c|c}
$\mathcal{B}$ & $M_{\mathcal{B}}$ [GeV] & $(\G/M)_{\mathcal{B}}$ \\
\hline
$\eta_c(1S)$	& $2.984$	 	& $1.1 \times 10^{-2}$ \\
$J/\psi(1S)$	& $3.097$		& $3.0 \times 10^{-5}$ \\
\hline
$\eta_b(1S)$	& $9.398$	 	& $1.2 \times 10^{-3}$ \\
$\Upsilon(1S)$	& $9.460$		& $5.7 \times 10^{-6}$
\end{tabular}
\end{table}

%----------------------------------------------------------------------------------------

In the charmonium case, the mass of the $1{}^1S_0$ bound state $\eta_c(1S)$ is $M_{\eta_c} \sim 3\GeV$.
The major difference in QCD is the existence of $N_F = 3$ light quarks below $\Lambda_\QCD$.
In this case, the $D$ mesons are the effective constituents of charmonium with $M_D \sim 1.85\GeV$ for the lightest $D$ meson;
therefore, the binding energy is approximately $0.6\TO0.7\GeV$, or about 20\% of the mass of $\eta_c(1S)$.
We then expect $m_X\sim400\TO450\GeV$.

To obtain the decay width $\G(\eta_X \to \g\g)$, we can perform a similar scaling with charmonium.
Using $\G_{\mathrm{tot}}/M$ for $\eta_c(1S)$ listed in Table~\ref{tab:qqmasses}
and $\Br(\eta_c\to\g\g) = (1.59 \pm 0.12) \times 10^{-4}$, both taken from Ref.~\cite{Agashe:2014kda},
we obtain
\be
\begin{split}
\frac{\G(\eta_X \to \g\g)}M &= \frac{\G(\eta_c \to \g\g)}{M_{\eta_c}} \left( \frac{Y_X}{Q_c} \right)^4 \\
&= 1.75 \times 10^{-6} \left( \frac{Y_X}{Q_c} \right)^4 \,,
\end{split}
\ee
where $Q_c = 2/3$ is the electromagnetic charge of the charm quark.
Substituting the above for $\G_{\mathrm{tot}}$ and with $\Br \sim 0.6$ in Eq.~(\ref{eqn:xs}), 
we thus see that the diphoton excess
can be accounted for $1.3 \lesssim Y_X \lesssim 1.6$, and for a wide range of
$\alpha_h$ (and equivalently, of $\Lambda_h$). 
This range is shown in the hashed region on the right-hand side of Fig.~\ref{fig:xsN3}.
Some representative values of these scales are sketched in Fig.~\ref{fig:largescales};
with the $X$ mass between $400\TO450\GeV$, $\Lambda_h$ is between
$100\TO150\GeV$, and the glueballs between $600\GeV$ and $1\TeV$,
reflecting the large uncertainties in the glueball mass.

The large binding energy in this case allows for the possibility of additional
bound states contributing to the signal.
We will elaborate on this possibility below.

%%%%%%%%%%%%%%%%%%%%%%%%%%%%%%%%%%%%%%%%%%%%%%%%%%%%%%%%%%%%%

\begin{table}[htb]
\centering
\caption{Summary of branching ratios to radiative decays for $J/\psi(1S)$ and $\Upsilon(1S)$, from Ref.~\cite{Agashe:2014kda}.
Decays in the last column proceed through a virtual photon.}
\label{tab:3S1BR}
\begin{tabular}{c|c|c}
$1{}^3S_1$ &$\Br(1{}^3S_1 \to (ggg,\gamma gg))$ & $\Br(1{}^3S_1 \to (\mathrm{had},\ell^+ \ell^-))$ \\
\hline
$J/\psi(1S)$	& $(0.641, 0.088)$ 	&$(0.135, 0.119)$ \\
$\Upsilon(1S)$	& $(0.817, 0.022)$ 	&$(\sim 0, 0.075)$
\end{tabular}
\end{table}

%%%%%%%%%%%%%%%%%%%%%%%%%%%%%%%%%%%%%%%%%%%%%%%%%%%
\section{Phenomenology: other LHC signals}
%%%%%%%%%%%%%%%%%%%%%%%%%%%%%%%%%%%%%%%%%%%%%%%%%

%----------------------------------------------------------------------------------------

\subsection{Additional bound states}

So far we considered threshold production of the $1^1S_0$ state $\eta_X$.
As mentioned above, 
additional bound states of different masses may contribute to the 
diphoton signal.
For example, in the charmonium system, $J/\psi(1S)$ decays to 
$\g + \eta_c(1S)$
with a branching ratio $0.017 \pm 0.004$.
The hyperfine mass splitting between the $\eta_c(1S)$ and the $J/\psi(1S)$ is approximately 113\,MeV,
or about 3.8\% of the $\eta_c(1S)$ mass.
Scaling this splitting to $M$ yields a mass for the $1^3S_1$ bound state, 
which we call $\U_X(1S)$ in analogy to bottomonium, that is $30\GeV$ above $M$.%
\footnote{This is consistent with the result expected from the confining potential, $\sim\Lambda^2/M$.} 
Therefore, the process $\U_X(1S)\to \eta_X \gamma$, which gives a relatively soft $\gamma$, can contribute to the diphoton signal.
This process also contributes to the $\eta_X$ diphoton resonance. 
In the Low-$\Lambda_h$ scenarios, the hyperfine splitting of the $1S$ state in the Coulomb approximation is given by
\begin{equation}
\Delta E_{\rm HF} \equiv E_b (1^3 S_1) - E_b (1^1 S_0) = \frac{1}{3} C^4 \overline{\alpha}_h^4 m_X \,,
\end{equation}
and can be $\mathcal{O}(10) \GeV$.

At threshold, $\eta_X$ is produced through VBF while $\Upsilon_X$ is produced through DY.%
\footnote{Production of more excited states (e.g., with $L > 0, n > 1$) is suppressed by their wavefunctions.}
For $N_F=0$, 
the possible $\eta_X$ and $\Upsilon_X$ decay channels are
\begin{equation}
 \begin{split}
   \eta_X &\to VV,~V G_h^*,~G_h G_h, \\
   \Upsilon_X &\to f\bar f,~VVV,~\eta_XV, ~G_h V,~G_h G_h^*,
 \end{split}
\end{equation}
where $V$ is a photon or a $Z$-boson, $f$ denotes a charged SM fermion, and $G_h^*$ is the excited $1^{+-}$ hidden glueball.

Among these, the process $\Upsilon_X \to l^+l^-$ is severely constrained by LHC dilepton resonance searches
~\cite{Aad:2014cka,Khachatryan:2014fba,CMS-PAS-EXO-15-005,ATLAS-CONF-2015-070},
where $l=e,\mu$.
These constraints imply
\begin{equation}
 \sigma(pp\to\eta_X\to\gamma\gamma)<\frac{1.3}{K_{q\bar q}}\fb \cdot Y_X^2
  \Bigl(\frac{C_{\gamma\gamma}}{78}\Bigr)\frac{\Br(\eta_X \to \gamma\gamma)}{\Br(\Upsilon_X \to f\bar f)} \,.
  \label{eq:upsilonbound}
\end{equation}
In the above, we have employed the bound $\sigma(pp\to\Upsilon_X\to l^+l^-)<1.2\fb$ at the $8\TeV$ LHC~\cite{Aad:2014cka}.\footnote{%
   This constraint assumes $\sigma(pp\to\Upsilon_X\to l^+l^-)\equiv\sigma(pp\to\Upsilon_X\to e^+e^-)=\sigma(pp\to\Upsilon_X\to \mu^+\mu^-)$.}
The relevant parton luminosities are taken from Ref.~\cite{Franceschini:2015kwy},
except for $C_{\gamma\gamma}$, which is extracted from Eq.~(\ref{eqn:xs}),
and we set $K_{\gamma\gamma}=1$.
In the Low-$\Lambda_h$ region, where we can calculate the different $\Upsilon_X$ decay rates, 
we find that, for $N_F=0$ and setting $K_{q \bar q}=1$, the resulting diphoton signal is smaller than 3~fb for $\oln{\alpha}_h \gtrsim 0.2$.
For larger values of $N_F$, higher values of $\oln{\alpha}_h$ would be allowed,
since $\Gamma(\Upsilon_X \to 3 g_h)$ would increase as $\alpha_h(m_X)$ increases.
In the High-$\Lambda_h$ case, the decay $\Upsilon_X \to G_h \gamma$ is open (see Fig.~\ref{fig:largescales});
however, to satisfy the constraint from $\Upsilon_X$ production, $\Gamma(\Upsilon_X \to G_h\gamma)$ 
must be enhanced by a factor of around 10 compared to the perturbative $\Gamma(\Upsilon_X \to g_hg_h\gamma)$.

Finally, radial excitations of $\eta_X$ may be produced as well. 
If these decay directly to diphotons (rather than to $\eta_X$ plus soft photons),
they could lead to a broad diphoton peak, of width 
$\lesssim E_b$~\cite{Kamenik:2016izk}.
In order to obtain the $\sim45$~GeV width 
favored by some analyses~\cite{ATLAS-CONF-2016-018}, 
we would require intermediate values for the couplings
in between the Low-$\Lambda_h$ and High-$\Lambda_h$ regions.

%----------------------------------------------------------------------------------------

\subsection{\secXXbar{} pair production above threshold}

Here the signatures crucially depend on the presence of light SU($N$) flavors.
For $N_F=0$, the models are essentially quirk models, with
the quirks carrying hypercharge only. 
The $X$--$\bar X$ pairs will form a string of length~\cite{Kang:2008ea}
\begin{equation}
 r\sim \frac{E}{\Lambda^2}\,,
\end{equation}
where $E$ is the quirk-pair energy.
In all our models, this string is microscopic:
\begin{equation}
 r \ll 1\,{\rm \mu m}\,.
\end{equation}
The quirk pair will then promptly annihilate into photons.
Note that the $X$--$\bar X$ pair is  produced with $L>0$,
and in principle, it could first relax to the bound state $\eta_X$,
which would subsequently decay to diphotons, contributing to the
diphoton resonance from $\eta_X$ threshold-production.
To lose angular momentum, the excited state must radiate photons
or glueballs. The latter process is kinematically suppressed
in the range of relatively large $\Lambda_h$ we consider,
so the only option is photon radiation.
If this energy loss process is efficient, 
the signal would be relatively soft photons from the relaxation process,
plus a diphoton resonance at $750\GeV$. 
This may greatly enhance the resonance, since DY production is larger than
photon fusion. 
However, with tight photon isolation cuts, these processes may be vetoed
because of the presence of additional photons.
The other possibility is that the radiation loss is not efficient, in which case $X$ and $\bar X$ 
annihilate without losing significant energy. 
This seems  more plausible in our models, 
given the tight binding of the quirk pair by the microscopic string.
The $X$--$\bar X$ pair can then go into either glueballs or photons, 
and the resulting photon invariant mass is simply the $X$--$\bar X$ invariant mass in this case.

Finally, the glueballs annihilate into SM particles, mainly photons,
through $X$ loops as mentioned above. These events would then have four or more
photons, with di- (or tri-)photon peaks at the glueball masses.

We now consider models with additional SU($N$) light flavors, with mass of order $\Lambda_h$.
As explained above, this is only viable in the Low-$\Lambda_h$ case.
The glueball lifetime is hardly affected, while the collider phenomenology as well as cosmology
are different since string breaking is no longer suppressed.
Consider an additional SM-singlet, SU($N$) fundamental field $S$,
which can be either a scalar or a vector-like fermion.
With no new couplings to SM fields, continuum $X$--$\bar X$ pair production
would be followed by hadronization into $X$--$\bar S$ mesons. 
We refer to these mesons as  $\xi_S$.
The lightest $\xi_S$ is charged and stable.
New couplings involving $X$, $S$, and SM fields must be introduced to mediate $\xi_S$ decays.
This restricts $Y_X$ to integer values.
The possible couplings are summarized in Table~\ref{tab:extracouplings}.

 \begin{table}[th]
  \caption{Lowest-order operators mediating $\xi_S$ decays.  
$\phi_S$ ($\chi_S$--$\bar\chi_S$) denotes a scalar (vector-like fermion) 
which is a SM singlet and SU($N$) (anti-)fundamental.
  SM flavor indices are omitted for simplicity.
  If $Y_X<0$, replace hidden sector fields by their conjugates.
  }
  \label{tab:extracouplings}
  \centering
 \begin{tabular}[t]{c|c|c}
        & scalar $\phi_S$ & vector-like fermion $\chi_S$ \\\hline
$Y_X=1$ &
     $\phi_S(\bar Xe^c)$&
         $\bar\chi_S Xll$
         \\
 $Y_X=2$ &
     $\phi_S^*Xu^cu^cu^c$, $\phi_S (\bar Xe^c) (ll)^*$&
          $\chi_S\bar Xe^ce^c$
\\
 $Y_X=3$ &
     $\phi_S\bar X e^ce^ce^c$ & none up to dim-7\\
\end{tabular}
 \end{table}
These decays provide various exotic signatures, with
$\xi_S$ pair production followed
by $\xi_S$ decay to $2\ell$, $3\ell$, or 3~jets, with $\ell=e, \mu$ or $\tau$. 
Events with $2\ell + \slashed{E}_T$ are possible too.
With the exception of the scalar coupling for $Y_X=1$, these couplings are
non-renormalizable, and therefore naturally small.
Thus, bound-state formation is still important, and $X$ particles indeed
hadronize before decaying.

Still, these operators can give sufficiently high rates to evade
the stringent constraints on long-lived (on detector scales) charged 
particles~\cite{Chatrchyan:2013oca,ATLAS:2014fka, CMS-PAS-EXO-15-010}%
\footnote{%
Charged particles with intermediate lifetimes, i.e., 
$1\,\mathrm{mm}\lesssim c \t \lesssim 1\,\mathrm{m}$, 
are also constrained at the LHC, though less severely~\cite{Aad:2013yna,Aad:2014gfa,CMS:2014hka,CMS:2014gxa,Aad:2015rba,Aad:2015qfa,Aaboud:2016dgf}.}
if the scale by which they are suppressed is $10\TeV$ or higher, depending on the operator.
Thus for example, for $Y_X=2$, even  the operator
 $(\lambda/M\w{high}^3)\phi_S(\bar Xe^c)(ll)^*$ gives
\begin{equation}
 c\tau\sim
  30\,\mathrm{\mu m}\cdot\frac1{\lambda^2}\,
  \left(\frac{M\w{high}}{10\TeV}\right)^6
  \left(\frac{1\GeV}{\Lambda_h}\right)^2
  \left(\frac{375\GeV}{M_{\xi_S}}\right)^5\,,
\end{equation}
while the analogous operator with $\chi_S$, allows for a higher $M\w{high}$
since it is dimension-4 only.

%----------------------------------------------------------------------------------------

\section{Conclusions}
In this paper, we considered the possibility that the observed
diphoton excess is due to a quarkonium-like bound state, $\eta_X$, of a hidden SU($N$),
with fermionic constituents carrying SM hypercharge only.
The production and decay of this bound state are controlled by two parameters:
$Y_X^2 \alpha$, which sets the coupling strength to photons,
and $C \alpha_h$, which controls the coupling to hidden gluons.
These scenarios lead to a variety of multi-photon signals, 
and possibly exotic decays to SM fermions.
Diphotons from photon-fusion production of $\eta_X$ are typically
accompanied by forward jets, with hadronic activity in the central region suppressed.

In large parts of the parameter space, production of the $J=1$ $\Upsilon_X$ bound state
leads to dilepton or dijet resonances close to 750~GeV.
Without additional hidden flavors or other dynamics, the bound Eq.~\eqref{eq:upsilonbound} excludes the parameter space above $\overline\alpha_h \gtrsim 0.2$.
For $\overline\alpha_h$ below 0.2, $\sigma(pp\to\Upsilon_X\to l^+l^-)$ at the $13\TeV$ LHC is between 1--3\,fb.

While we focused on a constituent fermion $X$ for concreteness, 
our results generalize trivially to the case of a scalar $X$. 
In particular, the blue curves of Figs.~\ref{fig:xsN3} and \ref{fig:xsN4} are barely modified, 
since the scalar production cross sections are down by a factor of 2 compared to the fermion case, 
but the signal strength scales at least as the fourth power of $Y_X$.
On the other hand, as mentioned in the Introduction, the production of $J=1$ bound states is suppressed in this case.

More generally, the relation between $\eta_X$ and $\Upsilon_X$ production depends on the details of the model.
For example, with $N_F$ flavors with masses somewhat below $m_X$, the renormalization group running of $\alpha_h$ between 
$m_X$ and the inverse Bohr radius is milder, resulting in larger $\alpha_h(m_X)$ for a given $\oln{\alpha}_h$.
This enhances the rates of both $\Upsilon_X$ and $\eta_X$ to hidden gluons, with the former increasing more sharply.
Second, in the presence of $S$--$X$ couplings to the SM, $X$ is unstable. 
If the decay width $\Gamma_X$ of $X$ is comparable to roughly half the bound state width,
bound state production is depleted. Since $\Gamma_{\eta_X} \gg \Gamma_{\Upsilon_X}$,
the $\Upsilon_X$ production cross section is substantially reduced if $\Gamma_X \sim \Gamma_{\Upsilon_X}/2$, 
whereas $\eta_X$ production is largely unaffected.
Third, a hidden $Z^\prime$ can mediate additional $\Upsilon_X$ decays to hidden flavors, 
while its effects on $\eta_X$ decays are milder.

Finally, $\eta_X$ production in association with additional photons or hidden gluons is quite generic in these models,
whether its source is higher resonances or continuum quirk-pair production for $N_F=0$ models.
Determining whether the diphoton resonance is accompanied by additional
softer photons or missing energy is therefore crucial.

%----------------------------------------------------------------------------------------

%----------------------------------------------------------------------------------------

\section{Acknowledgments}

%----------------------------------------------------------------------------------------

We thank Yevgeny Kats, Yuri Shirman, Emmanuel Stamou, Jure Zupan, and Jonathan Rosner for useful discussions. 
We also thank Yevgeny Kats and David Curtin for comments on an earlier version of this manuscript.
R.Z. thanks the Technion Particle Physics Center and Weizmann Institute of 
Science for hospitality while this work was initiated.
The research of S.I., G.L., and Y.S. is supported by the Israel Science Foundation (Grant No.~720/15), 
by the United States--Israel Binational Science Foundation (BSF) (Grant No.~2014397),
and by the ICORE Program of the Israel Planning and Budgeting Committee (Grant No. 1937/12).
The work of R.Z. was partially carried out in the ILP LABEX  (under reference ANR--10--LABX--63) 
and is supported by French state funds managed by the ANR within the Investissements d'Avenir programme under reference ANR--11--IDEX--0004--02.

%----------------------------------------------------------------------------------------
%	REFERENCES
%----------------------------------------------------------------------------------------

\bibliography{bound-states_diphoton_biblio}

\end{document}